Minor differences from final version published in *Nature* (511:567, July 31 2014)

# Misaligned Protoplanetary Disks in a Young Binary System

Eric L. N. Jensen[1] and Rachel Akeson[2]

[1] Dept. of Physics & Astronomy, Swarthmore College, 500 College Ave., Swarthmore, PA 19081, USA.

[2] NASA Exoplanet Science Institute, IPAC/Caltech, Pasadena, CA, 91125, USA.

**Many extrasolar planets follow orbits that differ from the nearly coplanar and circular orbits found in our solar system; orbits may be eccentric[1] or inclined with respect to the host star's equator[2,3], and the population of giant planets orbiting close to their host stars suggests significant orbital migration[4]. There is currently no consensus on what produces such orbits. Theoretical explanations often invoke interactions with a binary companion star on an orbit that is inclined relative to the planet's orbital plane[4,5]. Such mechanisms require significant mutual inclinations between planetary and binary star orbital planes. The protoplanetary disks in a few young binaries are misaligned[6-12], but these measurements are sensitive only to a small portion of the inner disk, and the three-dimensional misalignment of the bulk of the planet-forming disk mass has hitherto not been determined. Here we report that the protoplanetary disks in the young binary system HK Tau are misaligned by 60°–68°, so one or both disks are significantly inclined to the binary orbital plane. Our results demonstrate that the necessary conditions exist for misalignment-driven mechanisms to modify planetary orbits, and that these conditions are present at the time of planet formation, apparently due to the binary formation process.**

While the three-dimensional orbital orientation is not yet measurable for any of the known extrasolar planets, measuring the orientation of protoplanetary disks has the potential to provide information about planetary orbits during the planet formation process. Since these disks are hundreds of AU in diameter, they can be spatially resolved at the 120–160 pc distances of the nearest star-forming regions. (One AU is the average distance of the Earth from the Sun.) If the disks around both stars in a binary system can be shown to be misaligned, then it is clear that both cannot be aligned with the (usually undetermined) binary orbital plane. Indirect evidence of disk misalignment is provided by misaligned jets[9] and by polarimetry[13,14]. More directly, images of several young binary systems show that the disk around one star is nearly edge-on[6-8,12]. In some of these systems, infrared interferometry or imaging constrains the inclination of the disk around the other star, giving a lower limit on the degree of misalignment of the disks[8,12], though the position angle of the disks is uncertain and the direction of rotation is unknown. For systems with detectable millimeter-wavelength emission, measurement of Keplerian rotation in both disks in a binary system provides the opportunity to measure the full three-dimensional orientation of the disks' angular momentum.

One such system is HK Tau, a young binary system with a projected separation of 2.4 arcsec[15], which is 386 AU at the 161 parsec distance of this part of the Taurus clouds[16]. Age estimates for this system range from 1 to 4 Myr[17], clearly placing it in the age range at which planet formation is thought to occur. The southern, fainter star, HK Tau B, is surrounded by a disk that blocks the starlight; the disk can thus be clearly seen in scattered light images at near-infrared and visible wavelengths to be nearly edge-on[6,7,18]; statistical arguments suggest that the disk is unlikely to be completely aligned with the binary orbit[6,7]. The northern star, HK Tau A, has strong millimeter-wavelength continuum emission[19,20] showing that it too is surrounded by disk material, but since the disk does not block the starlight, the disk cannot be seen in scattered light due to the brightness of the star. The striking difference in their visible-light appearance shows that these two disks are not perfectly aligned, but the degree of misalignment has not previously been known because the molecular gas in the northern disk has not been resolved, and a modest inclination difference would be sufficient to explain the different scattered-light morphologies.

We observed HK Tau with the Atacama Large Millimeter Array (ALMA) at frequencies of 230.5 GHz and 345.8 GHz, covering continuum emission from dust and line emission from the carbon monoxide (CO) 2–1 and 3–2 rotational transitions, respectively (Methods). Both the northern and southern components of the binary are clearly detected in the continuum and the CO line emission. The CO maps (Fig. 1) show the clear signature of rotating disks around each star, with one side of the disk redshifted and the other side blueshifted. The orientations of the two disks are significantly different, with the northern disk axis elongated nearly north-south, roughly 45° from the elongation axis of the southern disk.

We used a Markov chain Monte Carlo (MCMC) analysis to fit disk models to our data to determine the three-dimensional spatial orientation of the disks (Methods). For HK Tau B, the disk orientation is well known from previous scattered-light imaging, and so we adopt from that work[18] an inclination $i$ = 85°±1° and position angle $PA$ = 42°. Though the disk inclination and position angle were previously known, our imaging of HK Tau B provides new spatial information since the direction of disk rotation, apparent in Fig. 1b, removes a 180° ambiguity in the disk's orientation. For what follows we adopt the convention that the position angle (PA) is measured east of north and that the quoted position angle is that of the redshifted edge of the disk. Our model fitting reproduces the individual velocity channel images well for both sources (Fig. 2), determining for the first time the position angle, inclination, and direction of rotation of the molecular gas disk in the northern source HK Tau A. The MCMC analysis gives *PA* =352°±3° and inclination $i$ = 43°±5° (Extended Data Fig. 1); all uncertainties are given as 68.3% credible intervals.

Measurement of the PA and inclination of both disks lets us determine the angle between the two disks' angular momentum vectors, with one ambiguity. Equal inclinations on either side of edge-on ($i$ = 90°) will appear identical unless it can be determined which edge of the disk is nearer to the observer, e.g. if high resolution imaging can determine that one edge of the disk is shadowed by a flared disk edge and the other is not. In the case of HK Tau B this orientation is known from scattered-light imaging, but it is still unknown for HK Tau A. Combining the observational constraints, we find that the angle between the two disks' angular momentum vectors is 60°±3° if both vectors point to the same side of the sky plane, or 68°±3° if they do not (Fig. 3).

The clear misalignment between the two disks has important implications for planet migration and orbital evolution, as well as theories of binary formation. While nothing in our observations constrains the orientation of the binary orbital plane, the fact that the two disks are misaligned with each other means that they cannot *both* be aligned with the binary orbital plane. At least one of the disks must be misaligned with the binary orbit by 30° (half the total misalignment) or more. The misalignment for one or both disks is likely greater than this, since this minimum misalignment only occurs for one specific orientation of the binary orbit. This misalignment means that planets formed from these disks will be subject to Kozai-Lidov oscillations[21-23] that may drive changes in their eccentricities and orbital inclinations, or the disks themselves may be driven into misalignment with the stars' rotation axes[5]. It is sometimes stated that only misalignments greater than the critical angle of 39.2° can cause Kozai-Lidov oscillations[21,23], but it has recently been shown that this is not strictly true if the body in the inner orbit is relatively massive and/or has an eccentric orbit[24]. In any case, it is quite likely that the inclination relative to the binary orbit exceeds this critical angle for one or both of the disks; only 1.6% of all possible binary orbits are inclined to both disks by less than 39.2° if the disks are misaligned by 60°.

This result is consistent with recent simulations of binary formation[25-27], which predict that disks will be misaligned with the binary orbit, especially in systems with orbital semimajor axes greater than 100 AU where dissipation mechanisms do not act quickly to align the disks with the orbit[25,28]. In earlier simulations of the formation of individual binary systems from isolated cloud cores, the level of misalignment depended on the choice of initial conditions[25]. However, more recent simulations[26,27] focus on the formation of entire clusters and thus do not presuppose specific initial conditions (or even a particular formation mechanism) for an individual binary[29]. In the cluster simulations of ref. 26, all binary systems with orbital semimajor axes greater than 30 AU have disks that are misaligned with each other, with a mean angle of $70° \pm 8°$. The misalignment we observe here is thus consistent with formation via turbulent fragmentation rather than disk instability[30].

While it remains to be seen how the protoplanetary disks in a statistical sample of young binary systems are oriented, it is suggestive that in the handful of systems where this measurement has been made, the misalignments are large. If this is a common outcome of the binary formation process, and especially if it extends to lower-mass binary companions (which may easily go undetected) as well, then perturbations by distant companions may account for many of the orbital properties that make the current sample of extrasolar planets so unlike our own solar system.

**Methods summary**

The CO 2–1 and 3–2 ALMA observations of HK Tau were calibrated using standard techniques. The antenna configuration yielded spatial resolutions (clean beam sizes) of 1.06” x 0.73” and 0.69” x 0.51” and spectral resolutions of 1.3 km $s^{-1}$ and 0.85 km $s^{-1}$ in the two bands. To determine the disk orientations, we calculated azimuthally symmetric, vertically isothermal parameterized disk models using a Monte Carlo radiation transfer code, and then sampled the model images at the same spatial frequencies and velocities as the observations to compare models to data in the *uv* plane. A Bayesian Markov Chain Monte Carlo analysis yielded posterior probability distributions for the disk parameters.

**Acknowledgements**

We thank Scott Schnee at NRAO for help in reducing the ALMA data, Steve Myers and Remy Indebetouw for assistance with ALMA data analysis, Lisa Prato for sharing data in advance of publication, and Meredith Hughes, David Cohen, Scott Gaudi, Lynne Steuerle Schofield, and Keivan Stassun for discussions. We are grateful to the referees, whose feedback improved the paper. ALMA is a partnership of ESO (representing its member states), NSF (USA) and NINS (Japan), together with NRC (Canada) and NSC and ASIAA (Taiwan), in cooperation with the Republic of Chile. The Joint ALMA Observatory is operated by ESO, AUI/NRAO and NAOJ. The National Radio Astronomy Observatory is a facility of the National Science Foundation operated under cooperative agreement by Associated Universities, Inc.

**Author Contributions**
ELNJ developed the disk modeling code, ran the models, and wrote most of the paper. RLA initiated the project, reduced the data, wrote the text on the observations, and commented on the paper draft.

**Author Information**
The authors declare no competing financial interests. Correspondence and requests for materials should be addressed to ejensen1@swarthmore.edu. This paper makes use of the following ALMA data: ADS/JAO.ALMA#2011.0.00150.S.

**Figures**

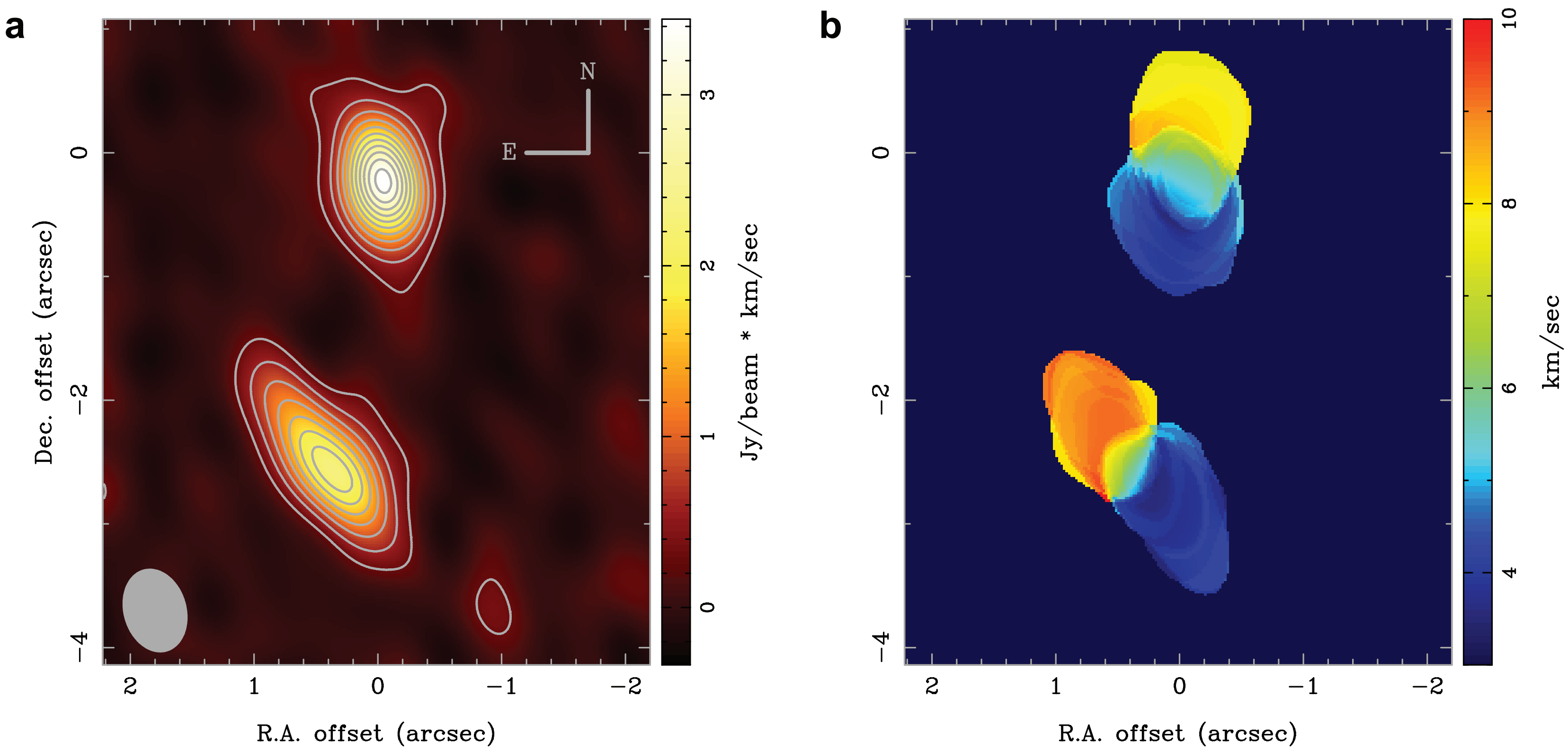

**Figure 1: Observations of the CO (3–2) line in the HK Tau binary system.** **a,** Integrated gas emission from each disk, with contours at steps of 0.3 Jy beam$^{-1}$ km s$^{-1}$, three times the RMS in the maps; the angular resolution of the observations is shown by the beam size in gray at lower left. **b,** Velocity-weighted emission, illustrating the rotation of both disks, and their misaligned orientations.

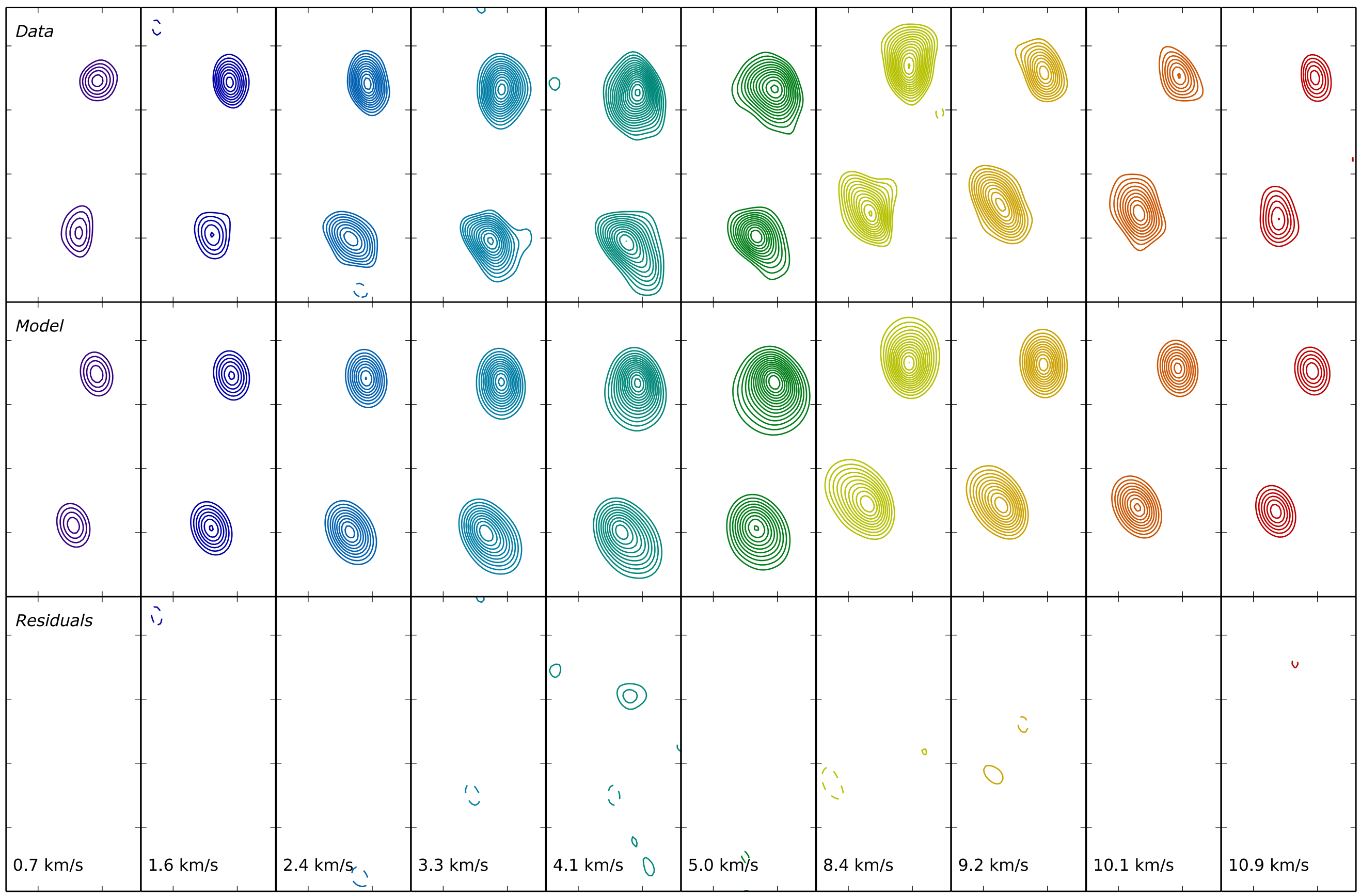

**Figure 2**: **Data, best-fit model, and data-model difference for the disks around HK Tau A and B.** Contours are in steps of 28 mJy, the RMS noise in the map, starting at three times the RMS. Negative contours are dashed. North is up and east is to the left, with tickmarks at 1-arcsecond intervals. Note that three channels near the line center of 6.1 km $s^{-1}$ are omitted from the figure and from calculating $\chi^2$ in the modeling due to absorption from the surrounding cloud.

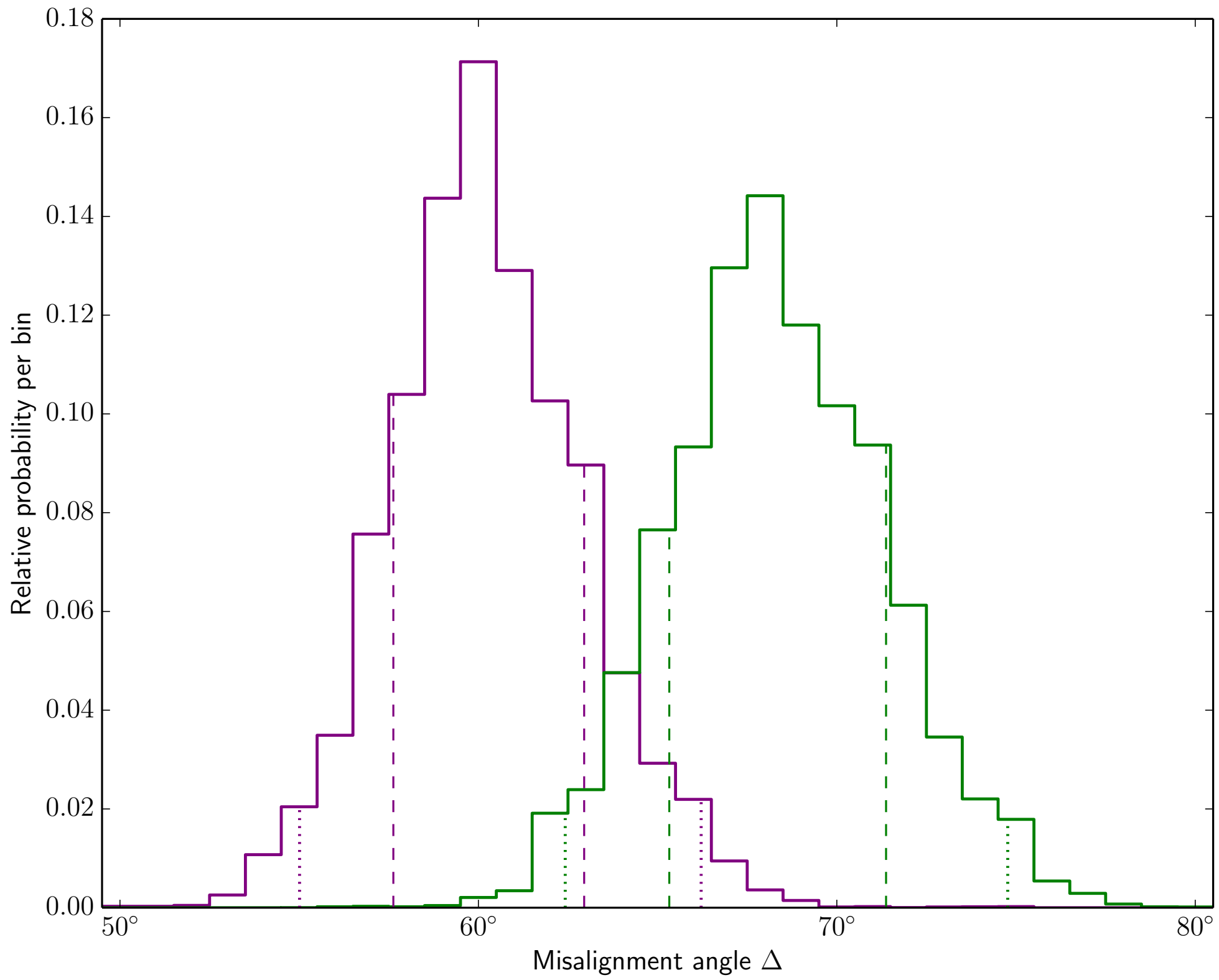

**Figure 3**: **Posterior probability distribution for the angle between the two disks' angular momentum vectors.** The purple histogram is for the case where both disks' vectors are on the same side of the sky plane; the green histogram is the case where they are on opposite sides of the sky plane. The 68.3% and 95.4% credible intervals are shown by dashed and dotted lines, respectively.

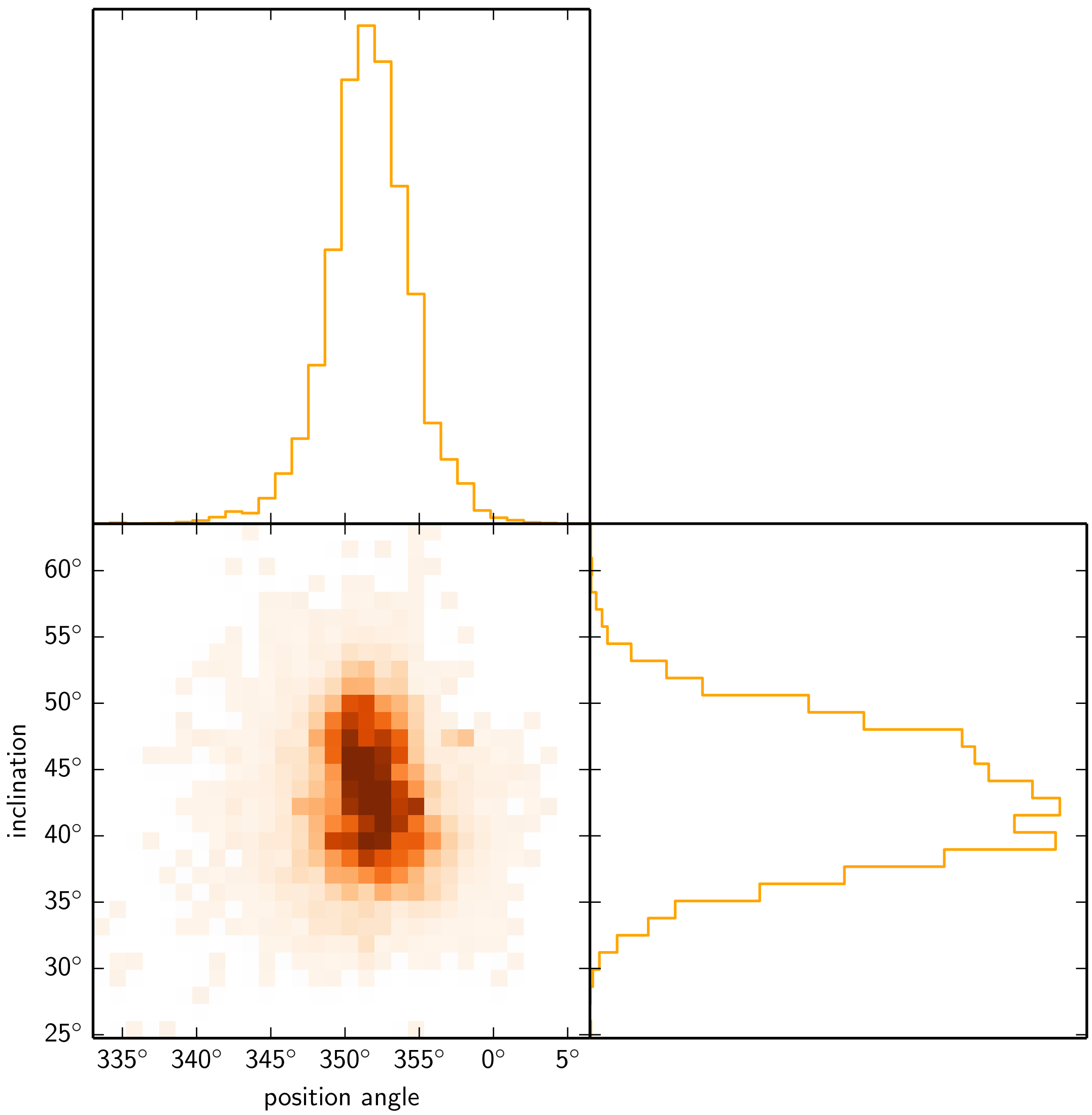

**Extended Data Figure 1**: Posterior probability distributions for the position angle and inclination of the disk around HK Tau A.

**Methods**

We observed HK Tau with the Atacama Large Millimeter Array (ALMA) as part of a survey of pre-main-sequence binaries in the Taurus-Auriga star-forming region[31]. Band 6 observations were taken on 17 Nov 2012 with 27 antennas and Band 7 observations on 16 Nov 2012 with 28 antennas. The correlator was configured with each of the four basebands covering a total bandwidth of 1.875 GHz with a channel spacing of 488 kHz. In Band 6, one of the correlator basebands was set to cover the CO (2–1) transition at 230.5 GHz, while in Band 7, one baseband covered CO (3–2) at 345.8 GHz. We took one observation of HK Tau at each band, bracketed by observations of the gain calibrator J051002+180041, which measures the phase and amplitude response as a function of time. We calibrated the data for each band separately using the *CASA* software and scripts provided by the NRAO ALMA center. The system temperature, water vapor phase corrections, and flagging were applied using the standard scripts. The amplitude and phase as a function of frequency were calibrated against J0423–013. The absolute flux calibration used Callisto and the 2012 flux models, which resulted in a zero spacing flux of 8.54 Jy at 230 GHz and 19.45 Jy at 345 GHz.

We generated continuum and CO images using the CLEAN task within *CASA*, with a robust beam weighting of –1.0. These settings resulted in a clean beam size of 1.06” x 0.73” in Band 6 and 0.69” x 0.51” in Band 7. The continuum flux of HK Tau is sufficient to provide a self-calibration reference and we applied a phase-only self-calibration using HK Tau as the reference. Given the short time on source, we averaged the continuum data to a single point in calculating the self-calibration corrections. The channel spacing, combined with Hanning smoothing in the correlator, provides a spectral resolution of 0.85 km $s^{-1}$ for the CO (3–2) line and 1.3 km $s^{-1}$ for the CO (2–1) line. The continuum emission is not strong enough to substantially affect the individual channels in the CO data and thus we did not subtract it.

The maps show clearly-detected CO emission centered at an LSR velocity of roughly 6.1 km $s^{-1}$. Examination of the individual channels of the CO data shows the presence of foreground absorption in the LSR velocity range of roughly 5–8 km $s^{-1}$, consistent with the absorption seen in the single-dish $^{13}$CO spectrum[32].

In order to quantify the disk properties, in particular the spatial orientation of each disk, we fit a series of models to the 345 GHz CO (3–2) data. Following many recent authors, we adopt a form for our disk model that is given by a self-similarity solution for circumstellar disks[33] and use the specific parameterization of ref 34.

While circumstellar disks in binary systems may be warped due to interactions with their stellar companions[28,35-37], the amount of warping is predicted to be largest for disks with aspect ratios less than 0.05. In contrast, the HK Tau B disk is relatively thick; with its measured scale height of 3.8 AU at a radius of 50 AU (ref. 18), the HK Tau B disk is predicted to have little or no warping. Assuming that the thickness of the HK Tau A disk is similar, warping should be of minimal importance for these disks, and thus we adopt an azimuthally symmetric disk model.

The gas density distribution in the model is azimuthally symmetric, and given by

$$\rho(r,z) = \frac{\Sigma(r)}{\sqrt{2\pi}H_p(r)} \exp\left[-\frac{1}{2}\left(\frac{z}{H_p(r)}\right)^2\right]$$

where $z$ is the vertical height above the disk midplane, and $\Sigma$ is the surface density distribution, given by

$$\Sigma(r) = \Sigma_c \left(\frac{r}{r_c}\right)^{-\gamma} \exp\left[-\left(\frac{r}{r_c}\right)^{2-\gamma}\right]$$

where $\Sigma_c$ is a constant such that the surface density at the characteristic radius $r_c$ is $\Sigma_c\, e^{-1}$ . $H_p$ is the pressure scale height, assumed to be in hydrostatic equilibrium and thus given by

$$H_p(r) = \left(\frac{kT(r)}{\mu m_H}\frac{r^3}{GM_*}\right)^{\frac{1}{2}}$$

where $T$ is the temperature, $k$ is Boltzmann's constant, $\mu$ is the mean molecular weight of the gas, $m_H$ is the mass of a hydrogen atom, and $M_*$ is the mass of the star. The disk is assumed to be vertically isothermal, and the radial temperature profile is assumed to be a power law and is normalized at 10 AU:

$$T(r) = T_{10}\left(\frac{r}{10\ AU}\right)^{-q}$$

Since the ambient radiation in the molecular cloud heats material even far from any star, we adopt a minimum temperature of 10 K, i.e. the power law above only applies out to the radius where $T(r)$ = 10 K, beyond which the temperature is constant at 10 K.

We assume that the dust and gas have the same temperature at a given radius, that the gas is in local thermodynamic equilibrium, that the gas-to-dust ratio by mass is 100, and that the number fraction of CO in the gas is $10^{-4}$. With these assumptions, there are six free parameters that characterize the disk emission and kinematics in the model: $M_{\rm disk}$, $r_{\rm c}$, $T_{10}$, $M_*$, $\gamma$ , and $q$. In addition, there are the two orientation parameters for the disk position angle *PA* and its inclination *i* to the line of sight. It is these latter two properties that are of primary interest to us for determining the disks' misalignment; the other six are varied in order to adequately reproduce the observed emission but we make no claim that they represent the true disk properties in detail, given the simplicity of the model and degeneracies between the parameters. We fix the positions of each component at the coordinates determined from fits to the velocity-integrated (first moment) maps of the CO emission, and we fix the line centers for both components at 6.1 km $s^{-1}$.

To find the distributions of parameter values that fit the data, we calculate a set of model disks using the Monte Carlo radiation transfer code *RADMC-3D* version 0.35 (ref. 38). The standard approach to comparing models to interferometric data is to transform the model images into the *uv* plane so that they can be compared directly with the data recorded by the interferometer, without the intervening, non-linear step of creating an image from the interferometric data. In the case of a binary system where both disks have strong emission, this presents an additional complication; while the two disks are cleanly separated in the image plane, their emission overlaps in the *uv* plane. Thus, it is necessary to compute models for both disks in order to compare models to data in the *uv* plane. This increases the number of free parameters for each step in the model-data comparison from 8 to 16, complicating the exploration of the parameter space.

To make this problem more tractable, we pursue a modeling strategy that rests on the assumption that the best-fit disk parameters for one star are uncorrelated with those of the other star, allowing us to fit for only 8 parameters at a time. As a preliminary step, we model the two disks in the HK Tau system individually. For each component of the binary, we use *RADMC-3D* with the model described above to create a single model disk, with images at different velocities across the CO (3–2) line that are separated by the velocity resolution of our observations. We then use the NRAO software *CASA* to sample the model image with the same *uv* coverage as our ALMA observations, and we create a CLEAN image in exactly the same way as we imaged our observations of HK Tau. The resultant model image is compared to a sub-image of our data with the same field of view, velocity channel spacing, and pixel scale, and we calculate $\chi^2$ between model and data. Using this image-plane modeling and the Markov Chain Monte Carlo (MCMC) analysis described in more detail below, we find the model parameters that provide the best fits for the A and B disks in the image plane.

Armed with these disk parameter estimates, we then proceed with the more robust *uv* plane modeling. To make the exploration of parameter space tractable, we vary parameters for only one disk at a time. In each model run, we hold constant the 8 parameters for one disk to values previously found to give a good fit, and vary only the 8 parameters for the other disk. We combine the two disk model images (one of which is always the same for a given run) into a single image with the disks centered at the known positions of HK Tau A and B. We then sample this model image with the same projected baselines used in the ALMA observations to generate model visibilities that can be compared directly with the data. We bin the data and models to 0.85 km $s^{-1}$ channels, the spectral resolution of the observations, and exclude the three channels near line center (LSR velocity range 5.4–7.9 km $s^{-1}$) where there is significant absorption from the cloud. We then calculate $\chi^2$ between model and data visibilities, with separate terms in the $\chi^2$ sum for the real and imaginary parts of each visibility point. The 10 channels shown in Figure 2 (spanning LSR velocities 0.3–5.4 and 7.9–11.3 km $s^{-1}$) are used in calculating $\chi^2$.

Because multiple combinations of the model parameters can provide almost equally good fits to the data, and because the parameter space is large, we use Markov Chain Monte Carlo (MCMC) to determine the posterior probability distribution of each parameter. As noted above, in each chain we vary only the 8 parameters for one of the disks. We use the Python code *emcee*[39], which implements an affine-invariant ensemble sampler[40]. For most parameters we use a flat prior probability, with the exception of inclination, where we use a sin *i* prior to account for the fact that randomly-distributed inclinations do not have equal probabilities of a given *i.* We evaluate the posterior probability of each model as $\exp(-\chi^2/2)$ times the prior probability. We ran several separate chains to explore a variety of starting positions for the disk's free parameters, and different fixed parameters for the other disk. In each chain, the ensemble had 30 "walkers" and ran for at least 500 steps. For each chain, we discarded the first 150 steps (4500 model evaluations) as "burn-in" so that the results would be independent of the starting positions chosen. Because the results from different chains were consistent with each other, we combined them to produce our final parameter estimates. Not including the burn-in steps, our final results for HK Tau A and HK Tau B are based on 66,000 and 30,000 model evaluations, respectively. As noted above, in the case of HK Tau B, the position angle and inclination are well known from scattered-light imaging, so for HK Tau B we adopt the *PA* and *i* values found from previous work in the analysis that follows, combined with our new measurements for HK Tau A.

The key quantity we are interested in determining is the angle $\Delta$ between the two disks' angular momentum vectors. It is related to the measured position angles and inclinations through spherical trigonometry by

$$\cos \Delta = \cos i_1 \cos i_2 + \sin i_1 \sin i_2 \cos \left(PA_1 - PA_2\right)$$

With both inclinations specified in the usual range of 0° to 90°, the above equation effectively assumes that both disks have their angular momentum vectors oriented on the same side of the plane of the sky. For the case where the two vectors are on opposite sides of the sky plane, one *i* above should be replaced with 180° – *i*. More specifically, we adopt the convention used in specifying the inclination of visual binary orbits[41], where $i < 90°$ corresponds to the case where the disk orbital motion is in the direction of increasing position angle, or equivalently where the disk's angular momentum vector is inclined by an angle 90° – *i* toward the observer relative to the sky plane. Thus, while our adopted convention for position angle (that of the redshifted edge of the disk) is the same as that typically adopted in previous work[42], our inclination convention differs.

By this convention, the inclination of the HK Tau B disk is 95° (since it is known from scattered light images that the northern face of the disk is tilted toward Earth), while the best-fit inclination of the HK Tau A disk could be either 43°±5° or 137°±5°. In practice, the two cases do not yield greatly differing values of $\Delta$ since HK Tau B is so close to edge-on.

In the near future, it may be possible to distinguish between these two inclinations for HK Tau A. A recently discovered Herbig Haro object, HH 678, lies 10 arcminutes west of HK Tau[43]. Its position angle of 267° with respect to HK Tau places it on a line that is nearly perpendicular to the HK Tau A disk, suggesting that it may be associated. If so, the sign of the radial velocity of the Herbig Haro object would break the inclination degeneracy for the HK Tau A disk.

We used fixed values of the orientation of the HK Tau B disk, and the values of *PA* and *i* for HK Tau A from our MCMC chains to find the posterior distribution for $\Delta$ of the two disks (Fig. 3). We take the median of the posterior distribution as the most probable value, and we find the values above and below the median that encompass 34.15% of probability in each direction in order to define the 68.3% credible interval (dashed lines); we similarly calculate the 95.4% credible interval (dotted lines). A plot of the posterior distributions of *PA* and *i* for HK Tau A (Extended Data Fig. 1) shows that they are uncorrelated, as expected.

Although our primary focus is the relative orientations of the disks, the modeling used here has the potential to determine other parameters of interest, in particular the stellar mass. Pre-main-sequence stellar mass measurements are of particular interest since they place valuable constraints on pre-main-sequence evolutionary models[44,45]. Unfortunately, due to our modest spatial resolution, coupled with the compact size of the HK Tau disks and the cloud absorption over several km $s^{-1}$ near the line center, we are unable to place tight constraints on the stellar masses. Our MCMC analysis yields $M_* = 0.6 \pm 0.1\ M_{\rm Sun}$ for HK Tau A and $M_* = 1.0 \pm 0.1\ M_{\rm Sun}$ for HK Tau B, where the quoted credible intervals do not take into account the uncertainty contribution from the distance to the HK Tau system. The HK Tau A mass is consistent with previous mass estimates from pre-main-sequence evolutionary tracks[17]. However, the HK Tau B mass is quite surprising. The published spectral types of HK Tau A and B are M1 and M2, respectively[13], and near-infrared high resolution spectra similarly yields spectral types of M0.5 and M1 for HK Tau A and B (L. Prato, in preparation). Given its later spectral type, and assumed coeval formation, HK Tau B should be less massive than HK Tau A. A possible resolution to the mass discrepancy would be if HK Tau B were itself a close binary. However, the near-infrared spectra show that the radial velocities of HK Tau A and B are the same to within 1 km $s^{-1}$, with no evidence of double lines in the spectra of either star (L. Prato, in preparation).

Thus, we suspect that our stellar mass estimate for HK Tau B may be inaccurate. It may be that our simple models do not adequately reproduce the vertical structure of the disk, which is likely to be much more important in modeling a nearly edge-on disk like HK Tau B than for one that is more face-on like HK Tau A. For example, ALMA science verification data of the disk around HD 163296 show that a vertical temperature gradient is necessary to reproduce the CO emission[46,47]. It is also possible that the uncertainty on the exact systemic velocity of the system (due to contamination from the molecular cloud) is a factor. Using a fixed systemic velocity parameter may introduce a small bias in the fit parameters, particularly the stellar mass. However, we see no structure in the residuals that would arise from using a systemic velocity far from the correct value.

We emphasize that the position angle and inclination for HK Tau B used in the analysis of disk misalignment were taken from previous scattered-light imaging, and thus modeling uncertainties for HK Tau B do not affect our main result here. Future ALMA data with better spatial resolution and using an isotopomer that is less sensitive to cloud absorption may help resolve the puzzle of HK Tau B’s stellar mass.